\documentclass[draft,journal,onecolumn]{IEEEtran}

\usepackage{cite}
\usepackage{graphicx}
\usepackage{amsmath}
\usepackage{amssymb}
\usepackage{amsthm}
\usepackage{amsfonts}
\usepackage{algorithmic}
\usepackage{algorithm}
\usepackage{mathrsfs}
\usepackage{cases}
\usepackage{mathtools}

\DeclareMathOperator{\RANK}{rank}

\DeclareMathOperator{\GRANK}{rank_G}
\DeclareMathOperator{\ERANK}{rank_E}

\newtheorem{definition}{Definition}
\newtheorem{theorem}{Theorem}
\newtheorem{lemma}{Lemma}

\newtheorem{construction}{Construction}
\newtheorem{remark}{Remark}

\newcommand{\OVERSET}[2]{\overset{\mathclap{\scriptscriptstyle{#1}}}{#2}}

\begin{document}

\title{Codes with Unequal Disjoint\\
    Local Erasure Correction Constraints
}

\author{
    Geonu~Kim,~\IEEEmembership{Student~Member,~IEEE,} and~Jungwoo~Lee,~\IEEEmembership{Senior~Member,~IEEE}%
    \thanks{
        The authors are with the Institute of New Media and Communications,
        Department of Electrical and Computer Engineering, Seoul National University, Seoul, 08826, Korea
        (e-mail: bdkim@wspl.snu.ac.kr; junglee@snu.ac.kr).
    }
}

\maketitle

\begin{abstract}
    Recently, locally repairable codes (LRCs) with local erasure correction constraints that are unequal and disjoint
    have been proposed.
    In this work, we study the same topic and provide some improved and additional results.
\end{abstract}

\IEEEpeerreviewmaketitle

\section{Introduction}

Locally repairable codes (LRCs) with \emph{unequal} or \emph{multiple} localities have been introduced
by \cite{Kadhe16ISIT} and \cite{Zeh16ISIT}, where localities are different among symbols.
This can be beneficial in the scenarios
where \emph{hot data} symbols require faster repair or reduced download latency \cite{Kadhe16ISIT}.

Both the results of \cite{Kadhe16ISIT} and \cite{Zeh16ISIT} rely on their own restrictive conditions.
In particular, \cite{Kadhe16ISIT} assumes the knowledge of the \emph{locality profile}
instead of the conventional locality, i.e., the \emph{locality requirement}.
On the other hand, \cite{Zeh16ISIT} assumes \emph{disjointness} of local repair groups of different locality.

It is natural to consider the extension of the $r$-locality to the $(r,\delta)$-locality.
This has been done in \cite{Kim17ARX} by the authors in the locality profile setting of \cite{Kadhe16ISIT}.
Similarly, \cite{Chen17ARX} extends the result of \cite{Zeh16ISIT}.

In this paper, we apply the techniques used in \cite{Kim17ARX} to the problem setting of \cite{Chen17ARX},
and obtain some improved and additional results.

The rest of the paper is organized as follows. In Section \ref{sect:preliminaries},
we review some important preliminaries.
The problem setting of unequal and disjoint $(r,\delta)$-localities is defined in Section \ref{sect:LRC-UDL}.
Section \ref{sect:bounds} presents both the dimension and minimum distance upper bounds,
and their tightness is shown by optimal code constructions in Section \ref{sect:construction}.
The concluding remarks are given in Section \ref{sect:conclusion}.

\section{Preliminaries} \label{sect:preliminaries}

\subsection{Notation}

We use the following notation.

\begin{enumerate}
    \item For an integer $i$, $ [i] = \{ 1, \ldots, i \} $.
    \item A vector of length $n$ is denoted by $ \mathbf{v} = (v_1,\ldots,v_n) $.
    \item A matrix of size $ k \times n $ is denoted by $ G = (g_{i,j})_{ i \in [k], j \in [n]} $.
    \item For sets $\mathcal{A}$ and $\mathcal{B}$, $ \mathcal{A} \sqcup \mathcal{B} $ denotes the disjoint union, i.e.,
        $ \mathcal{A} \cup \mathcal{B} $ with further implication that $ \mathcal{A} \cap \mathcal{B} = \emptyset $.
    \item For a symbol index set $ \mathcal{T} \subset [n] $ of a code $\mathscr{C}$ of length $n$,
        $\mathscr{C}\rvert_{\mathcal{T}}$ denotes the punctured code with support $\mathcal{T}$,
        and $G\rvert_{\mathcal{T}}$ is the corresponding generator matrix. Furthermore, we define
        $ \GRANK(\mathcal{T}) = \RANK(G\rvert_{\mathcal{T}}) $.
    \item For a symbol index set $ \mathcal{T} \subset [n] $ of a linear $[n,k]$ code $ \mathscr{C} $
        constructed via polynomial evaluation on an extension field $\mathbb{F}_{q^t}$,
        $\ERANK(\mathcal{T})$ denotes the rank of the evaluation points corresponding to $\mathscr{C}\rvert_{\mathcal{T}}$
        over the base field $\mathbb{F}_q$.
\end{enumerate}

\subsection{Minimum Distance}

The minimum distance of linear codes is well known to be characterized by the following lemma
\cite[Lem. A.1]{Kamath14TIT}, which is the basis of our minimum distance bounds.

\begin{lemma} \label{lem:dist-rank}
    For a symbol index set $ \mathcal{T} \subset [n]$ of a linear $[n,k,d]$ code such that $ \GRANK(\mathcal{T}) \leq k - 1 $, 
    we have 
    \begin{equation*}
        d \leq n - \lvert \mathcal{T} \rvert \text{,}
    \end{equation*}
    with equality if and only if $\mathcal{T}$ is of largest cardinality such that $ \GRANK(\mathcal{T}) = k - 1 $.
\end{lemma}

Below, we state a lemma (see also \cite{Kuijper14ARX}) based on Lemma \ref{lem:dist-rank}
that turns out to be more useful.
Note that this lemma can not be derived by simply substituting for $\lvert \mathcal{T} \rvert$ in Lemma \ref{lem:dist-rank}.

\begin{lemma} \label{lem:dist-red}
    For a symbol index set $ \mathcal{T} \subset [n]$ of a linear $[n,k,d]$ code such that $ \GRANK(\mathcal{T}) \leq k - 1 $,
    let $\gamma$ be the number of redundant symbols indexed by $\mathcal{T}$,
    i.e., $ \gamma = \lvert \mathcal{T} \rvert - \GRANK(\mathcal{T}) $.
    We have
    \begin{equation*}
        d \leq n - k + 1 - \gamma \text{.}
    \end{equation*}
\end{lemma} 

\begin{IEEEproof}
    Clearly, the set $\mathcal{T}$ can be enlarged to a set $\mathcal{T}'$ such that $ \GRANK(\mathcal{T}') = k - 1 $.
    Make another set $\mathcal{T}''$ by removing $\gamma$ redundant symbols from $\mathcal{T}'$.
    Note that $ \lvert \mathcal{T}'' \rvert \geq k - 1 $ since $ \GRANK(\mathcal{T}'') = k - 1 $.
    By applying Lemma \ref{lem:dist-rank} to the set $\mathcal{T}'$, we have
    \begin{align*}
        d & \leq n - \lvert \mathcal{T}' \rvert = n - \lvert \mathcal{T}'' \rvert - \gamma\\
        & \leq n - k + 1 - \gamma \text{.}
    \end{align*}
\end{IEEEproof}

As an immediate corollary to Lemma \ref{lem:dist-rank}, we also get the following lemma,
which is used when showing the optimal distance property of our code construction.
\begin{lemma} \label{lem:dist-rank:cor}
    For linear $[n,k,d]$ codes, if $ \GRANK(\mathcal{T}) = k $ for every symbol index set $ \mathcal{T} \subset [n] $
    such that $ \lvert \mathcal{T} \rvert = \tau $, we have
    \begin{equation*}
        d \geq n - \tau + 1 \text{.}
    \end{equation*}
\end{lemma}

\begin{remark} \label{rem:rank}
    In Lemma \ref{lem:dist-rank}, \ref{lem:dist-red}, and \ref{lem:dist-rank:cor},
    erasure correction is possible from $\mathcal{T}$ if and only if $ \GRANK(\mathcal{T}) = k $.
    Equivalently, erasure correction is not possible from $\mathcal{T}$ if and only if $ \GRANK(\mathcal{T}) \leq k - 1 $.
\end{remark}

\subsection{$(r,\delta)$-Locality}

A linear $[n,k,d]$ code $\mathscr{C}$ is said to have \emph{locality} $r$ (or $r$-locality)
if every symbol of $\mathscr{C}$ can be recovered with a linear combination of at most $r$ other symbols \cite{Gopalan12TIT}.
An equivalent description is that for each symbol index $ i \in [n] $, there exists a punctured code of $\mathscr{C}$ with
support containing $i$, length of at most $ r + 1 $ and distance of at least $2$.
We call such codes $r$-LRCs.
It has been shown in \cite{Gopalan12TIT} that the minimum Hamming distance $d$ of an $[n,k,d]$ $r$-LRC is upper bounded by
\begin{equation*}
    d \leq n - k + 2 - \left\lceil \frac{k}{r} \right\rceil \text{,}
\end{equation*}
which reduces to the well-known \emph{Singleton bound} if $ r \geq k $.
Various optimal code constructions achieving the equality in the minimum distance bound have been reported in the literature
\cite{Gopalan12TIT,Tamo13ISIT,Silberstein13ISIT,Song14JSAC,Tamo14TIT,Papailiopoulos14TIT,Goparaju14ISIT,Tamo15ISIT,Hao16ISIT}.

The notion of $r$-locality can be naturally extended to $(r,\delta)$-locality \cite{Prakash12ISIT}
to address the situation with multiple (local) node failures.
Note that $r$-locality corresponds to $(r,\delta=2)$-locality.

\begin{definition}[$(r,\delta)$-locality] \label{def:DLOC}
    A symbol with index $i\in[n]$ of a linear $[n,k,d]$ code $\mathscr{C}$ is said to have $(r,\delta)$-locality,
    if there exists a punctured code of $\mathscr{C}$ with support containing $i$,
    length of at most $ r + \delta - 1 $ and distance of at least $\delta$,
    i.e., there exists a symbol index set $ \mathcal{S}_i \subset [n] $ such that
    \begin{itemize}
        \item $ i \in \mathcal{S}_i $,
        \item $ \lvert \mathcal{S}_i \rvert \leq r + \delta - 1 $,
        \item $ d( \mathscr{C}\rvert_{\mathcal{S}_i} ) \geq \delta $.
    \end{itemize}
\end{definition}

\begin{remark} \label{rem:LCSB}
    By applying the Singleton bound to $\mathscr{C}\rvert_{\mathcal{S}_i}$ in Definition \ref{def:DLOC},
    we get $ \GRANK(\mathcal{S}_i) \leq r $ \textnormal{\cite{Kamath14TIT}}.
\end{remark}

Furthermore, $\mathscr{C}$ in the definition above is said to have $(r,\delta)$-locality
if every symbol of itself has $(r,\delta)$-locality, and is also called an $(r,\delta)$-LRC.
It is shown in \cite{Prakash12ISIT,Kamath14TIT} that the minimum distance of an $(r,\delta)$-LRC is upper bounded by
\begin{equation}
    d \leq n - k + 1 - \left( \left\lceil \frac{k}{r} \right\rceil - 1 \right) ( \delta - 1 ) \text{.} \label{eq:GBND}
\end{equation}
There are also several optimal code constructions in the literature
\cite{Prakash12ISIT,Kamath14TIT,Silberstein13ISIT,Tamo13ISIT,Tamo14TIT,Song14JSAC,Ernvall16TIT,Poellaenen16ISIT,Chen16ARX}
that achieve the equality in \eqref{eq:GBND}.

\subsection{$r$-LRC with Unequal Disjoint Localities}

While the locality of a code has been conventionally specified by a single parameter $r$,
recent works \cite{Zeh16ISIT,Kadhe16ISIT} have introduced the notion of \emph{unequal} (or \emph{multiple}) locality,
where the localities specified on different symbol sets are not equal.
Some restrictions that differ between \cite{Zeh16ISIT} and \cite{Kadhe16ISIT} are further imposed.
In this work, we build on the problem formulation of \cite{Zeh16ISIT},
which assumes some kind of \emph{disjointness}.
In particular, let $ [n] = \bigsqcup_{j=1}^{s} \mathcal{N}_j $ and $ \lvert\mathcal{N}_j\rvert = n_j $, $ j \in [s] $.
A linear $[n,k,d]$ code $\mathscr{C}$ is said to be have $\{(n_j,r_j)\}_{ j \in [s] }$-locality,
if every symbol with index $ i \in \mathcal{N}_j $, $ j \in [s] $,
is a linear combination of at most $r_j$ other symbols \emph{within} $\mathcal{N}_j$,
where $ r_1 \leq r_2 \leq \cdots \leq r_s $ without loss of generality.
The minimum distance of $\mathcal{C}$ is shown to be upper bounded by
\begin{equation}
    d \leq n - k + 2 - \sum_{j=1}^{s^*-1} \left\lceil \frac{n_j}{ r_j + 1 } \right\rceil -
        \left\lceil \frac{ k - \sum_{j=1}^{s^*-1} \left\lceil \frac{n_j}{ r_j + 1 } \right\rceil r_j }{r_{s^*}} \right\rceil
        \text{,} \label{eq:ZehBnd}
\end{equation}
where $ s^* = \max{\{ 0 \leq j \leq s-1 \mid
    \sum_{j'=1}^j \left\lceil \frac{n_{j'}}{ r_{j'} + 1 } \right\rceil r_{j'} < k - 1 \}} + 1 $.
Furthermore, a code construction based on shortening that is optimal with respect to this bound is provided,
thereby demonstrating the tightness of the bound.

On the other hand, \cite{Kadhe16ISIT} does not restrict the symbols composing the linear combination for
the symbol with index $ i \in \mathcal{N}_j $ to lie \emph{within} $\mathcal{N}_j$.
Instead, it is assumed that the locality for each symbol is specified in a minimum sense.
In other words, if a symbol is specified to have $r$-locality,
then it further means that it does not have $r'$-locality such that $ r' < r $,
i.e., the size of the minimal linear combination is $r$.
This minimum specification of symbol localities is defined by the notion of \emph{locality profile},
while the conventional specification is termed as the \emph{locality requirement}.
Under this restriction, \cite{Kadhe16ISIT} obtains a distance upper bound similar to \eqref{eq:ZehBnd},
which is also shown to be tight by an optimal code construction.

\subsection{Gabidulin Codes}

Our optimal code construction is an extension of the LRC construction based on Gabidulin codes
\cite{Silberstein13ISIT,Kadhe16ISIT}.
We thus give a brief introduction on Gabidulin codes, including some relevant properties.
Note that, due to the vector space structure of extension fields, an element in $\mathbb{F}_{q^t}$ can be
equivalently expressed as a vector of length $t$ over the base field $\mathbb{F}_q$, i.e., $\mathbb{F}_q^t$.

Gabidulin codes \cite{Gabidulin85} are an important class of maximum distance separable (MDS) codes%
\footnote{
    A generally more important property of Gabidulin codes is the maximum rank distance (MRD) property,
    but it is irrelevant in our derivations.
}.
Similar to Reed-Solomon (RS) and other algebraic codes, Gabidulin codes are constructed via polynomial evaluation.
However, both the data polynomials and the evaluation points are different from RS codes.
In particular, an $[n,k,d]_{q^t}$ Gabidulin code ($ t \geq n $) is constructed by encoding a message vector
$ \mathbf{a} = ( a_1, \ldots, a_k ) \in \mathbb{F}_{q^t}^k $ according to the following two steps.
\begin{enumerate}
    \item Construct a data polynomial $ f(x) = \sum_{i=1}^k a_i x^{q^{i-1}} $.
    \item Obtain a codeword by evaluating $f(x)$ at $n$ points $ \{ x_1, \ldots, x_n \} \subset \mathbb{F}_{q^t} $
        (or $\mathbb{F}_q^t$) that are linearly independent over $\mathbb{F}_q$, i.e.,
        $ \mathbf{c} = ( f(x_1), \ldots, f(x_n) ) \in \mathbb{F}_{q^t}^n $
        with $ \RANK(\{ x_1, \ldots, x_n \}) = n $.
\end{enumerate}
The data polynomial $f(x)$ belongs to a special class of polynomials
called \emph{linearized polynomials} \cite{Macwilliams77Book}.
The evaluation of a linearized polynomial over $\mathbb{F}_{q^t}$ is an $\mathbb{F}_q$-linear transformation.
In other words, for any $ a,b \in \mathbb{F}_q $ and $ x,y \in \mathbb{F}_{q^t} $, the following holds.
\begin{equation}
    f( ax + by ) = a f(x) + b f(y) \text{.} \label{eq:fqlin}
\end{equation}

The MDS property of Gabidulin codes can be shown by analyzing their erasure correction capability.
Specifically, the polynomial $f(\cdot)$, and therefore the underlying message vector $\mathbf{a}$,
can be recovered from evaluations on any $k$ points $\{ f(y_1), \ldots, f(y_k) \}$.
Since the $k$ evaluation points $\{ y_1, \ldots, y_k \}$ are linearly independent (over $\mathbb{F}_q$),
i.e., $ \RANK(\{ y_1, \ldots, y_k \}) = k $,
the use of the $\mathbb{F}_q$-linearity in \eqref{eq:fqlin} makes it possible to obtain evaluations at $q^k$ different points,
from which the polynomial $f(\cdot)$ of degree $q^{k-1}$ can be interpolated.
Therefore, erasure correction is possible from any $k$ symbols of the codeword.
Note that, the key property for erasure correction is that
the \emph{remaining rank} (of the evaluation points) corresponding to the remaining symbols is at least $k$.
In other words, $ n - k $ \emph{rank erasures} are tolerable.

In our code construction, we apply MDS encoding on chunks of a Gabidulin codeword
to equip the code with the desired local erasure correction property.
The following lemma, which is a special case of \cite[Lem. 9]{Rawat14TIT},
shows that symbols of such a code are also evaluations of the data polynomial $f(\cdot)$,
but the evaluation points generally differ from the original ones used in the Gabidulin codeword construction.

\begin{lemma} \label{lem:ERANK}
    For a vector $\mathbf{u}$ of length $k$ with elements being evaluations of a linearized polynomial $f(\cdot)$
    over $\mathbb{F}_{q^t}$, such that the evaluation points are linearly independent over $\mathbb{F}_q$,
    let $\mathbf{v}$ be the vector obtained by encoding $\mathbf{u}$ with an $[n,k]_q$ MDS code.
    Then any $s$ symbols of the codeword $\mathbf{v}$ correspond to evaluations of $f(\cdot)$ at $s$ points
    that lie in the subspace spanned by the original $k$ evaluation points (of $\mathbf{u}$).
    Furthermore the rank of the $s$ evaluation points is rank $\min(s,k)$,
    i.e., for an arbitrary set $ \mathcal{T} \subset [n] $ such that $ \lvert \mathcal{T} \rvert = s $, we have
    \begin{equation*}
        \ERANK(\mathcal{T}) = \min(s,k) \text{.}
    \end{equation*}
\end{lemma}

\begin{IEEEproof}
    We have $ \mathbf{u} = ( f(x_1), \ldots, f(x_k) ) \in \mathbb{F}_{q^t}^k $, $ \RANK(\{ x_1, \ldots, x_k \}) = k $,
    and $ \mathbf{v} = \mathbf{u} G $, where $G$ is the generator of the $[n,k]_q$ MDS code.
    Without loss of generality, denote the $s$ symbols in $\mathbf{v}$ as $\{ v_1, \ldots, v_s \}$, i.e.,
    $ \mathcal{T} = [s] $.
    We get
    \begin{equation*}
        ( v_1, \ldots, v_s ) = \mathbf{u} G\rvert_{[s]} = ( \sum_{i=1}^k g_{i,1}f(x_i), \ldots \sum_{i=1}^k g_{i,s}f(x_i) )
            \OVERSET{\eqref{eq:fqlin}}{=}
            ( f(\sum_{i=1}^k g_{i,1}x_i), \ldots f(\sum_{i=1}^k g_{i,s}x_i) ) \text{.}
    \end{equation*}
    Clearly, $v_j$ corresponds to an evaluation of $f(\cdot)$ at $ y_j = \sum_{i=1}^k g_{i,j}x_i $, for $ j \in [s] $.
    Furthermore, we get
    \begin{equation*}
        ( y_1, \ldots y_s ) = ( \sum_{i=1}^k g_{i,1}x_i, \ldots \sum_{i=1}^k g_{i,s}x_i ) =
            ( x_1, \ldots, x_k ) G\rvert_{[s]} \text{,}
    \end{equation*}
    and therefore
    \begin{equation*}
        \ERANK(\mathcal{T}) = \RANK(\{ y_1, \ldots, y_s \}) = \RANK(G\rvert_{[s]}) = \min(s,k) \text{.}
    \end{equation*}
\end{IEEEproof}

\section{$(r,\delta)$-LRC with Unequal Disjoint $(r,\delta)$-localities} \label{sect:LRC-UDL}

We generalize $r$-LRCs with unequal disjoint $r$-localities,
by defining $(r,\delta)$-LRCs with unequal disjoint $(r,\delta)$-localities,
which we call UD-$(r,\delta)$-LRCs in short (see also \cite[Def. 4]{Chen17ARX}).
We further define some auxiliary parameters to make various expressions more compact.

\begin{definition}[UD-$(r,\delta)$-LRC] \label{def:UDLRC}
    Let $ [n] = \bigsqcup_{j=1}^{s} \mathcal{N}_j $ and $ \lvert\mathcal{N}_j\rvert = n_j $, $ j \in [s] $.
    A linear $[n,k,d]$ code $\mathscr{C}$ is said to have $\{(n_j,r_j,\delta_j)\}_{ j \in [s] }$-locality
    if every symbol with index $ i \in \mathcal{N}_j $, $ j \in [s] $,
    has $(r_j,\delta_j)$-locality such that the corresponding punctured code satisfies $ \mathcal{S}_i \subset \mathcal{N}_j $.
    Furthermore, define
    \begin{itemize}
        \item integers $p_j$, $q_j$ such that $ n_j = p_j( r_j + \delta_j - 1 ) + q_j $
            and $ 0 \leq q_j \leq r_j + \delta_j - 2 $,
        \item $ \displaystyle m_j \triangleq \frac{n_j}{ r_j + \delta_j - 1 } = p_j + \frac{q_j}{ r_j + \delta_j - 1 } $,
        \item $ \displaystyle k_j \triangleq \begin{cases}
                \lfloor m_j \rfloor r_j & \text{if $ 0 \leq q_j \leq \delta_j - 2 $,} \\
                n_j - \lceil m_j \rceil ( \delta_j - 1 ) & \text{if $ \delta_j - 1 \leq q_j \leq r_j + \delta_j - 2 $.}
            \end{cases} $
    \end{itemize}
\end{definition}

Unlike the $r$-LRC case, we assume no order in the parameters $r_j$ and $\delta_j$, $ j \in [s] $, in the definition above.
However, more useful results are obtained in the special case where
the parameters $r_j$ and $\delta_j$ follows the two ordering conditions below (see also \cite[Def. 4]{Chen17ARX}).
Clearly, either condition alone can be assumed without loss of generality, but not both together.

\begin{definition}[Ordered $(r,\delta)$ condition] \label{def:ORDCON}
    An UD-$(r,\delta)$-LRC is said to satisfy the ordered $(r,\delta)$ condition if
    \begin{itemize}
        \item $ r_1 \leq r_2 \leq \cdots \leq r_s $,
        \item $ \delta_1 \geq \delta_2 \geq \cdots \geq \delta_s $.
    \end{itemize}
\end{definition}

\section{Upper Bounds} \label{sect:bounds}

\begin{algorithm} [t]
    \caption{Used in the Proof of Lemma \ref{lem:djrank} and \ref{lem:distbound}} \label{alg:Q}
    \begin{algorithmic}[1]
        \STATE Let $ \mathcal{Q}_0 = \emptyset $, $ l = 0 $
        \WHILE{$ \GRANK(\mathcal{Q}_l) < \GRANK( \mathcal{N}_j ) $} \label{alg:Q:while}
            \STATE Pick any $ i \in \mathcal{N}_j \setminus \mathcal{Q}_l $
                such that $ \GRANK( \mathcal{Q}_l \sqcup \{i\} ) > \GRANK(\mathcal{Q}_l) $ \label{alg:Q:pick}
            \STATE $ l = l + 1 $
            \STATE $ \mathcal{Q}_l = \mathcal{Q}_{l-1} \cup \mathcal{S}_i $
        \ENDWHILE
        \STATE $ L = l $
    \end{algorithmic}
\end{algorithm}

In this section, we first derive a dimension upper bound for UD-$(r,\delta)$-LRCs that does not depend on the minimum distance.
The Singleton-type minimum distance upper bound is provided next.
For the derivation of both of the bounds, we heavily rely on an algorithmic technique
that was originally proposed in \cite{Gopalan12TIT} and has been widely adapted in the literature
\cite{Kamath14TIT,Kadhe16ISIT,Zeh16ISIT}.
It is denoted as Algorithm \ref{alg:Q}.
The following remark and lemma (see also \cite[Lem. 5]{Kim17ARX} and the proof of \cite[Thm. 2.1]{Kamath14TIT})
describe the key properties of the algorithm.

\begin{remark} \label{rem:alg:Q}
    In Algorithm \ref{alg:Q}, $\mathcal{S}_i$ denotes the support of the punctured code
    by which the $i$th symbol has $(r_j,\delta_j)$-locality.
    Since $ \mathcal{S}_i \subset \mathcal{N}_j $, we have $ \mathcal{Q}_l \subset \mathcal{N}_j $
    and $ \GRANK(\mathcal{Q}_l) \leq \GRANK(\mathcal{N}_j) $.
    The condition in Step \ref{alg:Q:while} ensures that it is always possible to pick a suitable $i$ in Step \ref{alg:Q:pick}.
    The algorithm iterates until $ l = L $, where $ \GRANK(\mathcal{Q}_L) = \GRANK(\mathcal{N}_j) $.
\end{remark}

\begin{lemma} \label{lem:alg:Q}
    In Algorithm \ref{alg:Q}, we have, for $ l \in [L] $,
    \begin{enumerate}
        \item $ \GRANK(\mathcal{Q}_l) - \GRANK(\mathcal{Q}_{l-1}) \leq r_j $,
        \item $ \lvert \mathcal{Q}_l \rvert - \lvert \mathcal{Q}_{l-1} \rvert
            \geq \GRANK(\mathcal{Q}_l) - \GRANK(\mathcal{Q}_{l-1}) + \delta_j - 1 $,
        \item $ L \geq \left\lceil \GRANK(\mathcal{N}_j) \mathbin{/} r_j \right\rceil $.
    \end{enumerate}
\end{lemma}

\begin{IEEEproof}
    Since $ \mathcal{Q}_l = \mathcal{Q}_{l-1} \cup \mathcal{S}_i $, we have
    \begin{equation*}
        \GRANK(\mathcal{Q}_l) - \GRANK(\mathcal{Q}_{l-1}) \leq \GRANK(\mathcal{S}_i) \OVERSET{(a)}{\leq} r_j \text{,}
    \end{equation*}
    where (a) is due to Remark \ref{rem:LCSB}, and hence the first claim.

    For the second claim, first note that in the context of the punctured code with support $\mathcal{S}_i$,
    the symbols indexed by an arbitrary subset of $\mathcal{S}_i$ with the size of $ \delta_j - 1 $ are redundant
    since $ d(\mathscr{C}\rvert_{\mathcal{S}_i}) \geq \delta_j $.  
    We have $ \lvert \mathcal{Q}_l \rvert - \lvert \mathcal{Q}_{l-1} \rvert
        = \lvert \mathcal{Q}_l \setminus \mathcal{Q}_{l-1} \vert \geq \delta_j $,
    since otherwise we must have $ \GRANK(\mathcal{Q}_l) = \GRANK(\mathcal{Q}_{l-1}) $,
    due to the fact that $ \mathcal{Q}_l \setminus \mathcal{Q}_{l-1} \subset \mathcal{S}_i $.
    This is contradictory to the condition in Step \ref{alg:Q:pick}. 
    Now, out of the $ \lvert \mathcal{Q}_l \rvert - \lvert \mathcal{Q}_{l-1} \rvert \geq \delta_j $ incremental symbols
    in the set $\mathcal{Q}_l$, at least $ \delta_j - 1 $ symbols are redundant since they are already redundant
    in the context of $ \mathcal{S}_i \subset \mathcal{Q}_l $.
    Therefore, we get
    \begin{equation*}
        \GRANK(\mathcal{Q}_l) - \GRANK(\mathcal{Q}_{l-1})
            \leq \lvert \mathcal{Q}_l \rvert - \lvert \mathcal{Q}_{l-1} \rvert - ( \delta_j - 1 ) \text{.}
    \end{equation*}

    Finally, the first claim implies that
    \begin{equation*}
        L \geq \left\lceil \frac{\GRANK(\mathcal{Q}_L)}{r_j} \right\rceil \text{,}
    \end{equation*}
    and the last claim therefore directly follows from Remark \ref{rem:alg:Q}.

\end{IEEEproof}

We also require the following lemma (see also \cite[Lem. 6]{Kim17ARX}) on $\GRANK(\mathcal{N}_j)$,
both for the dimension and minimum distance upper bounds.

\begin{lemma} \label{lem:djrank}
    For UD-$(r,\delta)$-LRCs, we have
    \begin{equation*}
        \GRANK(\mathcal{N}_j) \leq k_j \text{,}
    \end{equation*}
    $ j \in [s] $.
\end{lemma}

\begin{IEEEproof}
    Considering the incremental symbols in the construction of $ Q_L \subset \mathcal{N}_j $ in Algorithm \ref{alg:Q},
    we obtain
    \begin{align}
        n_j & \geq \lvert \mathcal{Q}_L \rvert - \lvert \mathcal{Q}_0 \rvert
        = \sum_{l=1}^{L} ( \lvert \mathcal{Q}_l \rvert - \lvert \mathcal{Q}_{l-1} \rvert ) \notag \\
        & \OVERSET{(a)}{\geq} \sum_{l=1}^{L} ( \GRANK(\mathcal{Q}_l) - \GRANK(\mathcal{Q}_{l-1})) + L( \delta_j - 1 ) \notag \\
        & \OVERSET{(b)}{\geq}
            \GRANK(\mathcal{N}_j) + \left\lceil \frac{\GRANK(\mathcal{N}_j)}{r_j} \right\rceil ( \delta_j - 1 ) \text{,}
            \label{eq:lem:djrank}
    \end{align}
    where (a) comes from Lemma \ref{lem:alg:Q}-2), and (b) is due to Lemma \ref{lem:alg:Q}-3) and Remark \ref{rem:alg:Q}.

    For $ 0 \leq q_j \leq \delta_j - 2 $, suppose that $ \GRANK(\mathcal{N}_j) \geq p_j r_j + 1 $.
    It follows from \eqref{eq:lem:djrank} that
    \begin{align*}
        n_j & \geq p_j r_j + 1 + ( p_j + 1 )( \delta_j - 1 ) \\
        & = p_j ( r_j + \delta_j - 1 ) + \delta_j \\
        & > p_j ( r_j + \delta_j - 1 ) + q_j \\
        & = n_j \text{,}
    \end{align*}
    which is a contradiction. Therefore, we have
    \begin{equation*}
        \GRANK(\mathcal{N}_j) \leq p_j r_j = \lfloor m_j \rfloor r_j \text{.}
    \end{equation*}

    On the other hand, for $ \delta_j - 1 \leq q_j \leq r_j + \delta_j - 2 $,
    suppose that $ \GRANK(\mathcal{N}_j) \geq p_j r_j + q_j - ( \delta_j - 1 ) + 1 $,
    hence $ \GRANK(\mathcal{N}_j) \geq p_j r_j + 1 $.
    Again by \eqref{eq:lem:djrank}, we have
    \begin{align*}
        n_j & \geq p_j r_j + q_j - ( \delta_j - 1 ) + 1 + ( p_j + 1 )( \delta_j - 1 ) \\
        & = p_j ( r_j + \delta_j - 1 ) + q_j + 1 \\
        & > n_j \text{,}
    \end{align*}
    and therefore
    \begin{align*}
        \GRANK(\mathcal{N}_j) & \leq p_j r_j + q_j - ( \delta_j - 1 ) \\
        & = n_j - \lceil m_j \rceil ( \delta_j - 1 ) \text{.}
    \end{align*}
\end{IEEEproof}

The following theorem (see also \cite[Thm. 1]{Kim17ARX}) provides the dimension upper bound
as a simple corollary to Lemma \ref{lem:djrank}.

\begin{theorem}[Dimension upper bound] \label{thm:dimbound}
    The dimension of UD-$(r,\delta)$-LRCs is upper bounded by
    \begin{equation*}
        k \leq \sum_{j=1}^{s} k_j \text{.}
    \end{equation*}
\end{theorem}

\begin{IEEEproof}
    Clearly by Lemma \ref{lem:djrank}, we have $ k = \GRANK([n]) = \GRANK(\bigsqcup_{j=1}^{s} \mathcal{N}_j) \leq 
        \sum_{j=1}^{s} \GRANK(N_j) \leq \sum_{j=1}^{s} k_j $.
\end{IEEEproof}

Our minimum distance upper bound is based on the lemma below (see also \cite[Lem. 5]{Kim17ARX}),
where the $\GRANK(\mathcal{N}_j)$ terms appear in the expression.
They are subsequently eliminated in the theorem following the lemma (see also \cite[Thm. 2]{Kim17ARX}).

\begin{lemma} \label{lem:distbound}
    The minimum Hamming distance of UD-$(r,\delta)$-LRCs is upper bounded by
    \begin{equation*}
        d \leq n - k + 1 - \sum_{j=1}^{\sigma-1}(n_j - \GRANK(\mathcal{N}_j)) -
        \left(\left\lceil \frac{k - \sum_{j=1}^{\sigma-1}\GRANK(\mathcal{N}_j)}{r_{\sigma}} \right\rceil - 1 \right)
            (\delta_\sigma - 1) \text{,}
    \end{equation*}
    where
    \begin{equation*}
        \sigma = \min\{ j \in [s] \mid \sum_{j'=1}^{j} \GRANK(\mathcal{N}_{j'}) \geq k \} \text{.}
    \end{equation*}
\end{lemma}

\begin{IEEEproof}
    We build a set $ \mathcal{T} \subset [n] $ such that $ \GRANK(\mathcal{T}) \leq k - 1 $,
    and apply Lemma \ref{lem:dist-red} to obtain the distance upper bound.
    First, set $ j = \sigma $ in Algorithm \ref{alg:Q}.
    By Lemma \ref{lem:alg:Q}-3) and the definition of $\sigma$, we have
    \begin{equation*}
        L \geq \left\lceil \frac{\GRANK(\mathcal{N}_{\sigma})}{r_{\sigma}} \right\rceil \geq
        \left\lceil \frac{ k - \sum_{j=1}^{\sigma-1} \GRANK(\mathcal{N}_j) }{r_{\sigma}} \right\rceil \text{.}
    \end{equation*}
    For $\mathcal{Q}_l$ with
    \begin{equation*}
        l = \left\lfloor \frac{ k - 1 - \sum_{j=1}^{\sigma-1} \GRANK(\mathcal{N}_j) }{r_{\sigma}} \right\rfloor =
            \left\lceil \frac{ k - \sum_{j=1}^{\sigma-1} \GRANK(\mathcal{N}_j) }{r_{\sigma}} \right\rceil - 1 < L \text{,}
    \end{equation*}
    we get
    \begin{equation*}
        \GRANK(\mathcal{Q}_l) \OVERSET{(a)}{\leq} l \cdot r_\sigma \leq k - 1 - \sum_{j=1}^{\sigma-1} \GRANK(\mathcal{N}_j)
            \text{,}
    \end{equation*}
    where (a) is due to Lemma \ref{lem:alg:Q}-1).
    Therefore, letting $ \mathcal{T} = \left( \bigsqcup_{j=1}^{\sigma-1} \mathcal{N}_j \right) \sqcup \mathcal{Q}_l $, we have
    \begin{equation*}
        \GRANK(\mathcal{T}) \leq \sum_{j=1}^{\sigma-1} \GRANK(\mathcal{N}_j) + \GRANK(\mathcal{Q}_l) \leq k - 1 \text{.}
    \end{equation*}
    We conclude the proof by noting that the number of redundant symbols indexed by $\mathcal{T}$ is
    \begin{align*}
        \gamma & = \lvert \mathcal{T} \rvert - \GRANK(\mathcal{T}) \\ 
        & \geq \sum_{j=1}^{\sigma-1} n_j + \lvert \mathcal{Q}_l \rvert
            - \sum_{j=1}^{\sigma-1} \GRANK(\mathcal{N}_j) - \GRANK(\mathcal{Q}_l) \\
        & = \sum_{j=1}^{\sigma-1} ( n_j - \GRANK(\mathcal{N}_j) )
            + \sum_{l'=1}^{l} ( \lvert \mathcal{Q}_{l'} \rvert - \lvert \mathcal{Q}_{l'-1} \rvert )
            - \sum_{l'=1}^{l} ( \GRANK(\mathcal{Q}_{l'}) - \GRANK(\mathcal{Q}_{l'-1}) ) \\
        & \OVERSET{(a)}{\geq} \sum_{j=1}^{\sigma-1} ( n_j - \GRANK(\mathcal{N}_j) ) + l ( \delta_\sigma - 1 ) \\
        & = \sum_{j=1}^{\sigma-1} ( n_j - \GRANK(\mathcal{N}_j) ) +
            \left( \left\lceil \frac{ k - \sum_{j=1}^{\sigma-1} \GRANK(\mathcal{N}_j) }{r_{\sigma}} \right\rceil - 1 \right)
            ( \delta_\sigma - 1 ) \text{,}
    \end{align*}
\end{IEEEproof}

\begin{theorem}[Minimum distance upper bound] \label{thm:distbound}
    The minimum Hamming distance of UD-$(r,\delta)$-LRCs is upper bounded by
    \begin{equation*}
        d \leq n - k + 1 - \sum_{j=1}^{s^*-1} ( n_j - k_j )
            - \left( \left\lceil \frac{ k - \sum_{j=1}^{s^*-1} k_j }{r_{s^*}} \right\rceil - 1 \right)
            ( \delta_{s^*} - 1 ) \text{,}
    \end{equation*}
    where
    \begin{equation*}
        s^* = \min{\{ j \in [s] \mid \sum_{j'=1}^j k_{j'} \geq k \}} \text{.}
    \end{equation*}
\end{theorem}

\begin{IEEEproof}
    First note that $s^*$ is well defined due to Theorem \ref{thm:dimbound},
    and we have $ s^* \leq \sigma $ since $ \sum_{j=1}^\sigma k_j \geq \sum_{j=1}^\sigma \GRANK(\mathcal{N}_j) \geq k $.
    If $ s^* = \sigma $, it is easy to verify that the theorem holds
    by applying Lemma \ref{lem:djrank} on Lemma \ref{lem:distbound}.
    
    Otherwise, if $ s^* \leq \sigma - 1 $, we get
    \begin{align}
        d & \OVERSET{(a)}{\leq} n - k + 1 - \sum_{j=1}^{\sigma-1} ( n_j - \GRANK(\mathcal{N}_j) ) -
            \left( \left\lceil \frac{ k - \sum_{j=1}^{\sigma-1} \GRANK(\mathcal{N}_j) }{r_\sigma} \right\rceil - 1 \right)
            ( \delta_\sigma - 1 ) \notag \\
        & \OVERSET{(b)}{\leq} n - k + 1 - \sum_{j=1}^{s^*} ( n_j - \GRANK(\mathcal{N}_j) ) \notag \\
        & \OVERSET{(c)}{\leq} n - k + 1 - \sum_{j=1}^{s^*-1} ( n_j - k_j ) - ( n_{s^*} - k_{s^*} ) \text{,}
            \label{eq:thm:distbound}
    \end{align}
    where (a) is just Lemma \ref{lem:distbound}, (b) is obtained by removing some non-negative subtrahends,
    and (c) is due to Lemma \ref{lem:djrank}.

    Note that, if $ 0 \leq q_{s^*} \leq \delta_{s^*} - 2 $, we can write
    \begin{align}
        n_{s^*} - k_{s^*} & = n_{s^*} - \lfloor m_{s^*} \rfloor r_{s^*} \geq n_{s^*} - m_{s^*} r_{s^*} \notag \\
        & = m_{s^*} ( \delta_{s^*} - 1 ) \geq \lfloor m_{s^*} \rfloor( \delta_{s^*} - 1 ) \notag \\
        & = \frac{k_{s^*}}{r_{s^*}} ( \delta_{s^*} - 1 ) \text{.} \label{eq:thm:distbound:1}
    \end{align}
    Otherwise, if $ \delta_{s^*} - 1 \leq q_{s^*} \leq r_{s^*} + \delta_{s^*} - 2 $, again we get
    \begin{align}
        n_{s^*} - k_{s^*} &= \lceil m_{s^*} \rceil ( \delta_{s^*} - 1 ) \geq \frac{m_{s^*} r_{s^*}}{r_{s^*}}
            ( \delta_{s^*} - 1 ) \notag \\
        & = \frac{ n_{s^*} - m_{s^*} ( \delta_{s^*} - 1 ) }{r_{s^*}} ( \delta_{s^*} - 1 )
            \geq \frac{ n_{s^*} - \lceil m_{s^*} \rceil ( \delta_{s^*} - 1 ) }{r_{s^*}} ( \delta_{s^*} - 1 ) \notag \\
        & = \frac{k_{s^*}}{r_{s^*}} ( \delta_{s^*} - 1 ) \text{.} \label{eq:thm:distbound:2}
    \end{align}
    Furthermore, we have
    \begin{equation}
        \frac{k_{s^*}}{r_{s^*}} \geq \frac{ k - \sum_{j=1}^{s^*-1} k_j }{r_{s^*}} >
            \left\lceil \frac{ k - \sum_{j=1}^{s^*-1} k_j }{r_{s^*}} \right\rceil - 1 \text{.} \label{eq:thm:distbound:3}
    \end{equation}
    Therefore, substituting \eqref{eq:thm:distbound:1}, \eqref{eq:thm:distbound:2} and \eqref{eq:thm:distbound:3}
    into \eqref{eq:thm:distbound} completes the proof.
\end{IEEEproof}

Note that the bound given by Theorem \ref{thm:distbound} does not require
any ordering in the parameters $r_j$ and $\delta_j$, $ j \in [s] $.
Therefore, it is possible to obtain multiple bounds by permuting the index $j$.

The bound for the unequal disjoint $r$-locality case can be obtained by setting $ \delta_j = 2 $, $ j \in [s] $,
and the ordering in $r_j$ can be assumed without loss of generality.
Then, Theorem \ref{thm:distbound} results in a bound that is generally tighter than \eqref{eq:ZehBnd}.

\section{Optimal Code Construction} \label{sect:construction}

The minimum distance upper bound by Theorem \ref{thm:distbound} is of special interest,
in the case where the ordered $(r,\delta)$ condition by Definition \ref{def:ORDCON} holds.
In particular, we show the tightness of the bound by optimal code constructions for some parameter regime.
The code construction closely follows the Gabidulin-based LRC construction \cite{Silberstein13ISIT,Kadhe16ISIT}
(see also \cite[Construction 1]{Kim17ARX}).

\begin{construction}[Gabidulin-based UD-$(r,\delta)$-LRC] \label{cnstrct}
    For integers $m_j$, $ j \in [s] $, let $ n_j = m_j ( r_j + \delta_j - 1 ) $ and $ n = \sum_{j=1}^{s} n_j $.
    Let us also constrain the parameters to satisfy the condition
    $ k \leq n_{\mathbf{Gab}} \triangleq \sum_{j=1}^s m_j r_j \leq t $.
    Linear $[n,k,d]_{q^t}$ codes are constructed according to the following steps.
    \begin{enumerate}
        \item Precode $k$ information symbols using a Gabidulin code of length $n_{\mathbf{Gab}}$.
        \item Partition the Gabidulin codeword symbols into $ \sum_{j=1}^{r^*} m_j $ local groups,
            where each $m_j$ groups are of size $r_j$, $ j \in [s] $.
        \item Encode each local group of size $r_j$ using an $[r_j+\delta_j-1,r_j,\delta_j]_q$ MDS code.
    \end{enumerate}
\end{construction}

It is obvious by construction that a Gabidulin-based UD-$(r,\delta)$-LRC $\mathscr{C}$
has indeed $\{(n_j,r_j,\delta_j)\}_{j\in[s]}$-locality.
In particular, by choosing $\mathcal{S}_i$ as the support of the MDS local code corresponding to $ i \in \mathcal{N}_j $,
we have $ i \in S_i $ and $ \mathcal{S}_i = r_j + \delta_j - 1 $.
Furthermore, $ d(\mathscr{C}\rvert_{\mathcal{S}_i}) \geq \delta_j $ since
$\mathscr{C}\rvert_{\mathcal{S}_i}$ is a subcode of an $[r_j+\delta_j-1,r_j,\delta_j]_q$ MDS code. 

Note that, by having $ k = n_{\mathbf{Gab}} = \sum_{j=1}^s m_j r_j $ in the construction,
the equality in the dimension bound of UD-$(r,\delta)$-LRCs (Theorem \ref{thm:dimbound}) is achieved,
implying the tightness of the bound.

We require the following remark and lemma (see also \cite[Remark 5]{Kim17ARX} and \cite[Lem. 9]{Kim17ARX})
to analyze the minimum distance of the code by Construction \ref{cnstrct} satisfying the ordered $(r,\delta)$ condition,
which is shown to be optimal in the theorem following the lemma.

\begin{remark} \label{rem:dsum}
    Clearly, by Lemma \ref{lem:ERANK},
    the subspace generated by the evaluation points of the code of Construction \ref{cnstrct}
    is a direct sum of each subspace generated by the evaluation points corresponding to a single local group.
    Therefore, $\ERANK(\mathcal{T})$ of some set $ \mathcal{T} \subset [n] $ is the sum of each $\ERANK(\cdot)$
    computed separately on the points being in the same local group.
\end{remark}

\begin{lemma}\label{lem:rnkera}
    Let the parameters $\{(n_j,r_j,\delta_j)\}_{j\in[s]}$ in Construction \ref{cnstrct}
    satisfy the ordered $(r,\delta)$ condition (Definition \ref{def:ORDCON}).
    Suppose an ordered set $ \mathcal{L} = \{ \mathcal{G}_1, \ldots, \mathcal{G}_{\lvert \mathcal{L} \rvert } \} $
    such that $ \lvert \mathcal{L} \rvert = \sum_{j=1}^{s} m_j $,
    where each element of $\mathcal{L}$ is a symbol index set
    corresponding to the symbols of a distinct encoded local group in Construction \ref{cnstrct},
    and the order is according to the ordered $(r,\delta)$ condition.
    Elements of identical $(r,\delta)$ are ordered arbitrarily.
    Let denote an erasure pattern of $e$ erased symbols by the $ n - e $ remaining symbols.
    The index set of remaining symbols $ \mathcal{R}^* \subset [n] $, $ \lvert \mathcal{R}^* \rvert = n - e $,
    where the indices are taken greedily starting from the first element $\mathcal{G}_1$ of $\mathcal{L}$,
    corresponds to a worst case erasure pattern in terms of rank erasure (or remaining rank), i.e., we have
    \begin{equation*}
        \ERANK(\mathcal{R}) \geq \ERANK(\mathcal{R}^*) \text{,}
    \end{equation*}
    for any symbol index set $ \mathcal{R} \subset [n] $ such that $ \lvert \mathcal{R} \rvert = n - e $.
\end{lemma}

\begin{algorithm} [t]
    \caption{Used in the Proof of Lemma \ref{lem:rnkera}} \label{alg:R}
    \begin{algorithmic}[1]
        \WHILE{ $ \exists l_1,l_2 \in [\lvert \mathcal{L} \rvert] $, $ l_1 < l_2 $, such that
            $ \lvert \mathcal{R} \cap \mathcal{G}_{l_1} \rvert < \lvert \mathcal{G}_{l_1} \rvert $ and
            $ \lvert \mathcal{R} \cap \mathcal{G}_{l_2} \rvert > 0 $ }
            \STATE Construct $\Delta\mathcal{R}_1$ and $\Delta\mathcal{R}_2$ such that \\
                $ \Delta\mathcal{R}_1 \subset \mathcal{G}_{l_1} \setminus \mathcal{R} $, \\
                $ \Delta\mathcal{R}_2 \subset \mathcal{R} \cap \mathcal{G}_{l_2} $, and \\
                $ \lvert \Delta\mathcal{R}_1 \rvert = \lvert \Delta\mathcal{R}_2 \rvert =
                    \min( \lvert \mathcal{G}_{l_1} \setminus \mathcal{R} \rvert ,
                    \lvert \mathcal{R} \cap \mathcal{G}_{l_2} \rvert ) $
                \STATE $ \mathcal{R} = \mathcal{R} \sqcup \Delta\mathcal{R}_1 \setminus \Delta\mathcal{R}_2 $
                \label{alg:R:step}
        \ENDWHILE
    \end{algorithmic}
\end{algorithm}

\begin{IEEEproof}
    Any erasure pattern can be transformed into the claimed worst case pattern by repeatedly invoking Algorithm \ref{alg:R},
    since in Step \ref{alg:R:step} of the algorithm, symbols as many as possible in the local group $\mathcal{G}_{l_2}$
    are replaced with symbols in the local group $\mathcal{G}_{l_1}$, where $ l_1 < l_2 $.
    We show that that this replacement always results in a non-increasing \emph{remaining rank},
    making the claimed pattern worst indeed.

    First, observe that
    \begin{equation*}
        \mathcal{R} = \mathcal{R}_0 \sqcup \mathcal{R}_1 \sqcup \mathcal{R}_2 \text{,}
    \end{equation*}
    where
    \begin{align*}
        \mathcal{R}_0 & = \mathcal{R} \setminus ( \mathcal{R}_i \sqcup \mathcal{R}_j ) \text{,} \\
        \mathcal{R}_1 & = \mathcal{R} \cap \mathcal{G}_{l_1} \text{,} \\
        \mathcal{R}_2 & = \mathcal{R} \cap \mathcal{G}_{l_2} \text{.}
    \end{align*}
    We have
    \begin{equation}
        \ERANK(\mathcal{R}) \OVERSET{(a)}{=}
            \ERANK( \mathcal{R}_0 ) + \ERANK( \mathcal{R}_1 ) + \ERANK( \mathcal{R}_2 ) \text{,}
            \label{eq:lem:rnkera:r}
    \end{equation}
    where (a) is due to Remark \ref{rem:dsum}.
    Similarly, for
    \begin{align*}
        \mathcal{R}' & \triangleq \mathcal{R} \sqcup \Delta\mathcal{R}_1 \setminus \Delta\mathcal{R}_2 \\
        & = \mathcal{R}_0 \sqcup \mathcal{R}_1' \sqcup \mathcal{R}_2' \text{,}
    \end{align*}
    where
    \begin{align*}
        \mathcal{R}_1' & = \mathcal{R}' \cap \mathcal{G}_{l_1} = \mathcal{R}_1 \sqcup \Delta\mathcal{R}_1 \text{,} \\
        \mathcal{R}_2' & = \mathcal{R}' \cap \mathcal{G}_{l_2} = \mathcal{R}_2 \setminus \Delta\mathcal{R}_2 \text{,}
    \end{align*}
    we can write
    \begin{equation}
        \ERANK(\mathcal{R}') = \ERANK( \mathcal{R}_0 ) + \ERANK( \mathcal{R}_1' ) + \ERANK( \mathcal{R}_2' ) \text{.}
            \label{eq:lem:rnkera:rp}
    \end{equation}
    From \eqref{eq:lem:rnkera:r} and \eqref{eq:lem:rnkera:rp}, we have to show that
    \begin{equation}
        \ERANK(\mathcal{R}_1) + \ERANK(\mathcal{R}_2) \geq \ERANK(\mathcal{R}_1') + \ERANK({\mathcal{R}_2'}) \text{.}
            \label{eq:lem:rnkera}
    \end{equation}

    Let $\mathcal{G}_{l_i}$ be of $(r_{j_i},\delta_{j_i})$, $ i = 1,2 $.
    By the ordering of $\mathcal{L}$ and the ordered $(r,\delta)$ condition, we have
    $ r_{j_1} \leq r_{j_2} $ and $ \delta_{j_1} \geq \delta_{j_2} $.
    Note that by Lemma \ref{lem:ERANK}, we have
    \begin{align}
        \begin{split}
            \ERANK(\mathcal{R}_1) & = \min( \lvert \mathcal{R}_1 \rvert, r_{j_1} ) \text{,} \\
            \ERANK(\mathcal{R}_2) & = \min( \lvert \mathcal{R}_2 \rvert, r_{j_2} ) \text{,}
        \end{split} \label{eq:lem:rnkera:rnkr} \\
        \begin{split}
            \ERANK(\mathcal{R}_1') & = \min( \lvert \mathcal{R}_1 \rvert + \Delta, r_{j_1} ) \text{,} \\
            \ERANK(\mathcal{R}_2') & = \min( \lvert \mathcal{R}_2 \rvert - \Delta, r_{j_2} ) \text{,}
        \end{split} \label{eq:lem:rnkera:rnkrp}
    \end{align}
    where
    \begin{align}
        \Delta & = \lvert \Delta\mathcal{R}_1 \rvert = \lvert \Delta\mathcal{R}_2 \rvert =
            \min( \lvert \mathcal{G}_{l_1} \setminus \mathcal{R} \rvert , \lvert \mathcal{R} \cap \mathcal{G}_{l_2} \rvert )
            \notag \\
        & = \min( \lvert \mathcal{G}_{l_1} \rvert - \lvert \mathcal{R}_1 \rvert, \lvert \mathcal{R}_2 \rvert )
            \text{.} \label{eq:lem:rnkera:D}
    \end{align}
    We only provide the proof for the case where $ \lvert \mathcal{R}_1 \rvert \leq r_{j_1} $ and
    $ \lvert \mathcal{R}_2 \rvert > r_{j_2} $,
    since it is easy to verify that \eqref{eq:lem:rnkera} holds in other cases.
    From \eqref{eq:lem:rnkera:rnkr}, we have
    \begin{equation*}
        \ERANK(\mathcal{R}_1) + \ERANK(\mathcal{R}_2) = \lvert \mathcal{R}_1 \rvert + r_{j_2} \text{.}
    \end{equation*}
    If $ \lvert \mathcal{G}_{l_2} \rvert - \lvert \mathcal{R}_1 \rvert \geq \lvert \mathcal{R}_2 \rvert $,
    from \eqref{eq:lem:rnkera:D} and \eqref{eq:lem:rnkera:rnkrp}, we get
    \begin{align*}
        \ERANK(\mathcal{R}_1') + \ERANK(\mathcal{R}_2') & \leq r_{j_1} \leq r_{j_2} \\
        & \leq \lvert \mathcal{R}_1 \rvert + r_{j_2} \text{,}
    \end{align*}
    and therefore \eqref{eq:lem:rnkera}.
    Otherwise, \eqref{eq:lem:rnkera} again holds since
    \begin{align*}
        \ERANK(\mathcal{R}_1') + \ERANK(\mathcal{R}_2') & \leq
            r_{j_1} + \lvert \mathcal{R}_1 \rvert + \lvert \mathcal{R}_2 \rvert - \lvert \mathcal{G}_{l_1} \rvert \\
        & = \lvert \mathcal{R}_1 \rvert + \lvert \mathcal{R}_2 \rvert - ( \delta_{j_1} - 1 ) \\
        & \leq \lvert \mathcal{R}_1 \rvert + \lvert \mathcal{G}_{l_2} \rvert - ( \delta_{j_2} - 1 ) \\
        & = \lvert \mathcal{R}_1 \rvert + r_{j_2} \text{.}
    \end{align*}
\end{IEEEproof}

\begin{theorem}[Optimality of Gabidulin-based UD-$(r,\delta)$-LRCs with ordered $(r,\delta)$] \label{thm:optOL}
    Gabidulin-based UD-$(r,\delta)$-LRCs (Construction \ref{cnstrct})
    satisfying the ordered $(r,\delta)$ condition (Definition \ref{def:ORDCON})
    are distance optimal with respect to the distance upper bound for UD-$(r,\delta)$-LRCs (Theorem \ref{thm:distbound}).
\end{theorem}

\begin{IEEEproof}
    We derive a lower bound on the minimum distance of the code, which equals the upper bound of Theorem \ref{thm:distbound}.
    In particular, we show that erasure correction is possible from an arbitrary symbol set with the cardinality of
    \begin{equation*}
        \tau = k + \sum_{j=1}^{s^*-1} m_j ( \delta_j - 1 ) +
            \left( \left\lceil \frac{ k - \sum_{j=1}^{s^*-1} m_j r_j }{r_{s^*}} \right\rceil - 1 \right)
            ( \delta_{s^*} - 1 ) \text{,}
    \end{equation*}
    where $s^*$ is given by Theorem \ref{thm:distbound}.
    Applying Lemma \ref{lem:dist-rank:cor} with Remark \ref{rem:rank} gives the desired lower bound.
    
    Let integers $P$ and $Q$ such that
    \begin{equation}
        k - 1 - \sum_{j=1}^{s^*-1} m_j r_j = P r_{s^*} + Q \geq 0 \label{eq:thm:optOL:pqdef}
    \end{equation}
    and $ 0 \leq Q \leq r_{s^*} - 1 $.
    Consider an arbitrary symbol index set $ \mathcal{T} \subset [n] $ of cardinality
    \begin{equation}
        \lvert \mathcal{T} \rvert = \sum_{j=1}^{s^*-1} n_j + P ( r_{s^*} + \delta_{s^*} - 1 ) + Q + 1 \text{.}
            \label{eq:thm:optOL:Tcard}
    \end{equation}
    Let $\mathcal{T}^*$ be the greedily chosen set of Lemma \ref{lem:rnkera}
    such that $ \lvert \mathcal{T}^* \rvert = \lvert \mathcal{T} \rvert $,
    which consists of all the symbols in the local groups of $(r_j,\delta_j)$, $j \in [s^*-1] $,
    $P$ local groups of $(r_{s^*},\delta_{s^*})$,
    and some $ Q + 1 $ symbols in an additional local group of $(r_{s^*},\delta_{s^*})$.
    This composition is valid since
    \begin{align*}
        P & \OVERSET{\eqref{eq:thm:optOL:pqdef}}{=} \frac{ k - \sum_{j=1}^{s^*-1} m_j r_j }{r_{s^*}}
            - \frac{ 1 + Q }{r_{s^*}} \\
        & \OVERSET{(a)}{\leq} \frac{ m_{s^*} r_{s^*} }{r_{s^*}} - \frac{ 1 + Q }{r_{s^*}} \\
        & < m_{s^*} \text{,}
    \end{align*}
    where (a) comes from the definition of $s^*$.
    We have
    \begin{align*}
        \ERANK(\mathcal{T}) & \OVERSET{(a)}{\geq} \ERANK(\mathcal{T}^*) \\
        & \OVERSET{(b)}{=} \sum_{j=1}^{s^*-1} m_j r_j + P r_{s^*} + Q + 1 \\
        & \OVERSET{\eqref{eq:thm:optOL:pqdef}}{=} k \text{,}
    \end{align*}
    where (a) is Lemma \ref{lem:rnkera}, and (b) is due to Lemma \ref{lem:ERANK} and Remark \ref{rem:dsum},
    hence erasure correction is possible from $\mathcal{T}$.

    The proof is complete by noting that
    substituting \eqref{eq:thm:optOL:pqdef} into \eqref{eq:thm:optOL:Tcard} yields
    \begin{equation*}
        \lvert \mathcal{T} \rvert = k + \sum_{j=1}^{s^*-1} m_j ( \delta_j - 1 ) + P( \delta_{s^*} - 1 ) \text{,}
    \end{equation*}
    which is equal to $\tau$ since
    \begin{equation*}
        P \OVERSET{\eqref{eq:thm:optOL:pqdef}}{=}
            \left\lfloor \frac{ k - \sum_{j=1}^{s^*-1} m_j r_j - 1 }{r_{s^*}} \right\rfloor =
            \left\lceil \frac{ k - \sum_{j=1}^{s^*-1} m_j r_j }{r_{s^*}} \right\rceil - 1 \text{.}
    \end{equation*}
\end{IEEEproof}

\section{Conclusion} \label{sect:conclusion}

In this work, we have investigated codes with unequal disjoint $(r,\delta)$-locality. 
A minimum distance upper bound has been obtained, which is shown to be tight
by a Gabidulin-based optimal code construction under the ordered $(r,\delta)$ condition.
A tight dimension upper bound characterizing the feasible rate region is also derived.

\bibliographystyle{IEEEtran}
\bibliography{IEEEabrv,draft}

\begin{thebibliography}{10}
\providecommand{\url}[1]{#1}
\csname url@samestyle\endcsname
\providecommand{\newblock}{\relax}
\providecommand{\bibinfo}[2]{#2}
\providecommand{\BIBentrySTDinterwordspacing}{\spaceskip=0pt\relax}
\providecommand{\BIBentryALTinterwordstretchfactor}{4}
\providecommand{\BIBentryALTinterwordspacing}{\spaceskip=\fontdimen2\font plus
\BIBentryALTinterwordstretchfactor\fontdimen3\font minus
  \fontdimen4\font\relax}
\providecommand{\BIBforeignlanguage}[2]{{%
\expandafter\ifx\csname l@#1\endcsname\relax
\typeout{** WARNING: IEEEtran.bst: No hyphenation pattern has been}%
\typeout{** loaded for the language `#1'. Using the pattern for}%
\typeout{** the default language instead.}%
\else
\language=\csname l@#1\endcsname
\fi
#2}}
\providecommand{\BIBdecl}{\relax}
\BIBdecl

\bibitem{Kadhe16ISIT}
S.~Kadhe and A.~Sprintson, ``Codes with unequal locality,'' in \emph{2016 IEEE
  International Symposium on Information Theory (ISIT)}, July 2016, pp.
  435--439.

\bibitem{Zeh16ISIT}
A.~Zeh and E.~Yaakobi, ``Bounds and constructions of codes with multiple
  localities,'' in \emph{2016 IEEE International Symposium on Information
  Theory (ISIT)}, July 2016, pp. 640--644.

\bibitem{Kim17ARX}
\BIBentryALTinterwordspacing
G.~Kim and J.~Lee, ``Locally repairable codes with unequal locality
  requirements,'' \emph{CoRR}, vol. abs/1701.07340, 2017. [Online]. Available:
  \url{http://arxiv.org/abs/1701.07340}
\BIBentrySTDinterwordspacing

\bibitem{Chen17ARX}
\BIBentryALTinterwordspacing
B.~Chen, S.~Xia, and J.~Hao, ``Locally repairable codes with multiple
  (r\({}_{\mbox{i}}\), {\(\delta\)}\({}_{\mbox{i}}\))-localities,''
  \emph{CoRR}, vol. abs/1702.05741, 2017. [Online]. Available:
  \url{http://arxiv.org/abs/1702.05741}
\BIBentrySTDinterwordspacing

\bibitem{Kamath14TIT}
G.~M. Kamath, N.~Prakash, V.~Lalitha, and P.~V. Kumar, ``Codes with local
  regeneration and erasure correction,'' \emph{IEEE Transactions on Information
  Theory}, vol.~60, no.~8, pp. 4637--4660, Aug 2014.

\bibitem{Kuijper14ARX}
\BIBentryALTinterwordspacing
M.~Kuijper and D.~Napp, ``Erasure codes with simplex locality,'' \emph{CoRR},
  vol. abs/1403.2779, 2014. [Online]. Available:
  \url{http://arxiv.org/abs/1403.2779}
\BIBentrySTDinterwordspacing

\bibitem{Gopalan12TIT}
P.~Gopalan, C.~Huang, H.~Simitci, and S.~Yekhanin, ``On the locality of
  codeword symbols,'' \emph{IEEE Transactions on Information Theory}, vol.~58,
  no.~11, pp. 6925--6934, Nov 2012.

\bibitem{Tamo13ISIT}
I.~Tamo, D.~S. Papailiopoulos, and A.~G. Dimakis, ``Optimal locally repairable
  codes and connections to matroid theory,'' in \emph{Information Theory
  Proceedings (ISIT), 2013 IEEE International Symposium on}, July 2013, pp.
  1814--1818.

\bibitem{Silberstein13ISIT}
N.~Silberstein, A.~S. Rawat, O.~O. Koyluoglu, and S.~Vishwanath, ``Optimal
  locally repairable codes via rank-metric codes,'' in \emph{Information Theory
  Proceedings (ISIT), 2013 IEEE International Symposium on}, July 2013, pp.
  1819--1823.

\bibitem{Song14JSAC}
W.~Song, S.~H. Dau, C.~Yuen, and T.~J. Li, ``Optimal locally repairable linear
  codes,'' \emph{IEEE Journal on Selected Areas in Communications}, vol.~32,
  no.~5, pp. 1019--1036, May 2014.

\bibitem{Tamo14TIT}
I.~Tamo and A.~Barg, ``A family of optimal locally recoverable codes,''
  \emph{IEEE Transactions on Information Theory}, vol.~60, no.~8, pp.
  4661--4676, Aug 2014.

\bibitem{Papailiopoulos14TIT}
D.~S. Papailiopoulos and A.~G. Dimakis, ``Locally repairable codes,''
  \emph{IEEE Transactions on Information Theory}, vol.~60, no.~10, pp.
  5843--5855, Oct 2014.

\bibitem{Goparaju14ISIT}
S.~Goparaju and R.~Calderbank, ``Binary cyclic codes that are locally
  repairable,'' in \emph{2014 IEEE International Symposium on Information
  Theory}, June 2014, pp. 676--680.

\bibitem{Tamo15ISIT}
I.~Tamo, A.~Barg, S.~Goparaju, and R.~Calderbank, ``Cyclic {LRC} codes and
  their subfield subcodes,'' in \emph{2015 IEEE International Symposium on
  Information Theory (ISIT)}, June 2015, pp. 1262--1266.

\bibitem{Hao16ISIT}
J.~Hao, S.~T. Xia, and B.~Chen, ``Some results on optimal locally repairable
  codes,'' in \emph{2016 IEEE International Symposium on Information Theory
  (ISIT)}, July 2016, pp. 440--444.

\bibitem{Prakash12ISIT}
N.~Prakash, G.~M. Kamath, V.~Lalitha, and P.~V. Kumar, ``Optimal linear codes
  with a local-error-correction property,'' in \emph{Information Theory
  Proceedings (ISIT), 2012 IEEE International Symposium on}, July 2012, pp.
  2776--2780.

\bibitem{Ernvall16TIT}
T.~Ernvall, T.~Westerback, R.~Freij-Hollanti, and C.~Hollanti, ``Constructions
  and properties of linear locally repairable codes,'' \emph{IEEE Transactions
  on Information Theory}, vol.~62, no.~3, pp. 1129--1143, March 2016.

\bibitem{Poellaenen16ISIT}
A.~Pollanen, T.~Westerback, R.~Freij-Hollanti, and C.~Hollanti, ``Bounds on the
  maximal minimum distance of linear locally repairable codes,'' in \emph{2016
  IEEE International Symposium on Information Theory (ISIT)}, July 2016, pp.
  1586--1590.

\bibitem{Chen16ARX}
\BIBentryALTinterwordspacing
B.~Chen, S.-T. Xia, J.~Hao, and F.-W. Fu, ``Constructions of optimal cyclic
  $(r,\delta)$ locally repairable codes,'' \emph{CoRR}, vol. abs/1609.01136,
  2016. [Online]. Available: \url{http://arxiv.org/abs/1609.01136}
\BIBentrySTDinterwordspacing

\bibitem{Gabidulin85}
{\`E}.~M. Gabidulin, ``Theory of codes with maximum rank distance,''
  \emph{Problemy Peredachi Informatsii}, vol.~21, no.~1, pp. 3--16, 1985.

\bibitem{Macwilliams77Book}
F.~MacWilliams and N.~Sloane, \emph{The Theory of Error Correcting
  Codes}.\hskip 1em plus 0.5em minus 0.4em\relax North-Holland Publishing
  Company, 1977.

\bibitem{Rawat14TIT}
A.~S. Rawat, O.~O. Koyluoglu, N.~Silberstein, and S.~Vishwanath, ``Optimal
  locally repairable and secure codes for distributed storage systems,''
  \emph{IEEE Transactions on Information Theory}, vol.~60, no.~1, pp. 212--236,
  Jan 2014.

\end{thebibliography}

\end{document}